# $d^5$-off-centering induced ferroelectric and magnetoelectric correlations in trirutile-Fe$_2$TeO$_6$

P. Pal[1], S. D. Kaushik[2], Shalini Badola[3], S. Kuila[4], Parasmani Rajput[5], Surajit Saha[3], P. N. Vishwakarma[1], A. K. Singh[1]

[1]*Department of Physics and Astronomy, National Institute of Technology, Rourkela-769008, India*

[2]*UGC-DAE Consortium for Scientific Research, Bhabha Atomic Research Centre Mumbai-400085, India*

[3]*Department of Physics, Indian Institute of Science Education and Research, Bhopal-462066, India*

[4]*Department of Physics, Indian Institute of Technology Bombay, Mumbai-400076, India*

[5]*Atomic & Molecular Physics Division, Bhabha Atomic Research Center, Trombay, Mumbai- 400085, India*

**Abstract:**

We present the rare existence of $d^5$ off-centering, weak ferroelectric polarization and demonstrate its correlation with observed magnetoelectric (ME) properties in the G type ($T_N$~210 K) antiferromagnet Fe$_2$TeO$_6$ (FTO) compound. The origin of ferroelectricity (FE) is associated with both lattice and asymmetric electron density distribution around the ion cores. ME coupling is observed in magnetic field-dependent polarization, ME voltage, and magnetostrain measurements. Short-range magnetic ordering due to intrabilayer dimeric exchange coupling via the double oxygen bridged Fe-O1-Fe pathway is proposed to play a dominating role to exhibit the negative nonlinear magnetic field dependent ME behavior at 300 K. Interbilayer exchange via Fe-O2-Fe pathways dominantly determines the hysteretic nonlinear magnetic field dependent ME response below $T_N$. The observed nonlinear ME coupling signifies magnetoelasticity as manifested in the temperature and magnetic field-dependent strain measurement. Hence the rare existence of ferroelectricity and magnetoelectric coupling by $d^5$ ion is presented in FTO.

## I. Introduction:

Intertwining different degrees of freedom such as spin, lattice, and charge degrees of freedom can generate a plethora of novel physical properties in some material called quantum materials [1, 2]. The coupling of these three degrees of freedom, precisely in magnetoelectric (ME) multiferroic materials, is captivating due to its prospects toward technological revolution [3-5]. The simultaneous coexistence of coupled ferroelectricity and different magnetic ferro order in these materials generates the novel properties like quantum criticality on the emergence of electrocaloric properties, memory effect on domain wall conductivity, ferrotoroidicity, spontaneous ordering of magnetic vortices, *etc.* [6, 7]. The origin of ferroelectricity and ME coupling mechanisms are rich and diverse in nature as explored in the last decade [8]. Natural "contraindication" generally suggests ferroelectricity originated due to $d^0$ ions and magnetism originating from $d^n$ ion [9, 10]. Several mechanisms have been understood overcoming the "contraindication" in the past decade through the ferroelectricty due to $d^n$ electronic structure and magnetism due to $d^0$ electronic structure apparently forbidden. The origin of ferroelectricity connected with $d^n$ ions in multiferroic material diversified the field of multiferroic material research. Due to weak off-centering of dn ions locally, searching for ferroelectricity can be a new way out of the "contraindication" [11].

Tetragonal Trirutile *(P42/mnm)* AFM oxide of $A_2^{+3}B^{+6}O_6^{-2}$ (A=Fe, Cr; B=Te, W) type are collinear ME material [12]. This class of material exhibits ME properties based on the symmetry possessed by the magnetic atom [13]. The emergence of FE on the application of the magnetic field is reported in $Cr_2WO_6$ [14]. The origin of ME coupling below the AFM transition ($T_N$) was explained by the formation of ME domain due to the applied magnetic field [14]. ME coupling below $T_N$ was reported in $Fe_2TeO_6$ (FTO) by magnetic susceptibility measurements after poling under an electric field [15]. It was proposed that the two-ion mechanisms could explain the ME coupling. However, it was not elaborated on the microscopic cross-correlation between lattice and spin near $T_N$ was not elaborated. The presence of short-range magnetic order was proposed as a signature in broad maximum in the susceptibility above $T_N$ [16, 17]. Recently, the existence of spin dimers and their role in defining quantum criticality, suppressing magnetic moment, and the existence of the magnetic correlation above $T_N$ are reported in isostructural $Cr_2TeO_6$ and $Cr_2WO_6$ via inelastic neutron scattering [18]. Our group had also reported FE and ME coupling in FTO at 300 K [19, 20]. The Mössbauer study on this material reveals that the coordination polyhedra of $Fe^{3+}$ are distorted locally though structurally it was not reported [16]. The lack of microscopic understanding, ferroelectric, and ME properties motivated us to investigate the crystal, magnetic, and its correlation with ME coupling in FTO.

In this work, detailed crystal and magnetic structure are studied by neutron diffraction to investigate the origin of ferroelectricity and ME coupling present in the material. The structural information is obtained from the neutron diffraction analyzed by Ellipsoid analysis to investigate the coordination polyhedra of $Fe^{3+}$ and $Te^{6+}$. Local information about the charge, *p-d* hybridization, and coordination polyhedra are studied by X-ray absorption near edge (XANE) and extended X-ray absorption fine structure (EXAFS). The role of this local distortion in $Fe^{3+}$ polyhedra on the electron density distribution is inspected by calculating the electron density distribution by MEM analysis using the structural information obtained by the Rietveld refinement of synchrotron X-ray diffraction data. The observed electron density distribution around different cations is used to calculate the electronic contribution to polarization. Finally, we reinforce the existence of finite $d^5$ off-centering of $Fe^{3+}$ with asymmetric electron density distribution to induce FE and ME coupling in FTO with experimental evidence.

## II.     Experimental techniques:

The polycrystalline sample $Fe_2TeO_6$ (FTO) is prepared using high purity (~99.9%; Sigma Aldrich), $Fe_2O_3$, and $TeO_2$ via conventional solid-state reaction routes. All the ingredient oxides are mixed with proper stoichiometric ratio and grounded thoroughly for 2 h. This grounded mixture is calcined at 750°C and followed by reground and sintering at 750°C for 2 h. The ferroelectric and ME voltage measurements are performed on the sintered (750°C) pellet. The pellets are coated with silver for the electrode.  The synthesized sample is subjected to a temperature-dependent neutron powder diffraction (ND) experiment with a neutron beam with a wavelength of 1.48 Å. The diffraction experiment is carried out using the multi-position sensitive detector based focusing crystal diffractometer established by UGC-DAE CSR Mumbai Centre at the National Facility for Neutron Beam Research (NFNBR) Dhruva reactor, Mumbai (India). The X-ray ($\lambda=0.6304Å$) diffraction measurement is performed for charge density distribution analysis at the BL-11 of Indus-2 synchrotron source at the Raja Ramanna Centre for Advanced Technology (RRCAT)) Indore, India. Fe K-edge X-ray absorption spectroscopy (XAS) for both X-ray absorption near-edge spectroscopy (XANES) and Extended X-ray absorption fine structure (EXAFS) study are done at the BL-9 of Indus-2 synchrotron source at RRCAT, Indore, India. The remnant FE polarization is measured using a ferroelectric loop tracer (M/s. Precision LC-II of Radiant Inc.) under different fixed magnetic fields (0 T-1.2 T) at room temperature. ME voltage is measured under applied DC magnetic field modulated with AC field at different fixed temperatures from 125 K to 300 K. The modulating field is applied by a Helmholtz coil fed by AC through a lock-in-amplifier (M/s. Stanford research system make SR830). The ME voltage is measured using the lock-in –amplifier (SR830). The strain measurement is done by a commercially available standard strain gauge and the gauge resistance is measured by a multimeter (Keithley 2000). The DC magnetic field is applied by the M/s. GMW make electromagnet powered by M/s. KEPCO make power supply.

### III. Results and discussions:

**A. Neutron diffraction and Ellipsoid analysis of coordination polyhedra:**

The Rietveld analyses of the ND data are performed by considering tetragonal trirutile crystal structure with $P4_2/mnm$ (space group no.136). For the analysis of the magnetic structure, the basis representation vectors for magnetic Fe atom were calculated using *BasIrreps* under Fullprof [21] using propagation vector K as [000], consistent with the previous study [12]. The corresponding tetragonal structure (Fig. 1(a)) consists of four layers in the unit cell. The uppermost and lowermost layer consists of four $Te^{6+}$ polyhedra at the corner, and in between the two layers, there are another two layers of $Fe^{3+}$ polyhedra. There is an alternate arrangement of the $Fe^{3+}$ and $Te^{6+}$ ions at the body center positions of the trirutile subcompartment in between any two layers. It can be seen clearly ( Fig.1) that upon decreasing the temperature, new peaks at 20.8˚, 22.4˚, and 43.4˚ start developing below 250 K, indicating AFM ordering. AFM arrangement (Fig. 2(a)) of the magnetic moments along the *c-axis* is obtained by the magnetic refinement of the diffraction pattern at 3 K. The Fe moment at 3 K is obtained as ~ 4.35 $\mu_B$, which is ~13% lower than the $Fe^{3+}$ moment in high spin (5/2) state. From the analysis, we further determine the temperature-dependent cell parameters and magnetic moment (inset), as shown in Fig. 2(b). The $Fe^{3+}$ polyhedra share edges and also with the $Te^{6+}$ polyhedra on either side. $Te^{6+}$

polyhedra follow $D_{4h}$ (Trans) symmetry at all the observed temperatures. In the case of $Fe^{3+}$ polyhedra, in-plane (equatorial) unequal Fe-O and equal Fe-O bond lengths on the out of plane (apical) coordination apparently suggest $D_{4h}$ (Cis) symmetry. The side panel (lower) of Fig. 2(a) shows the distorted polyhedra with different bond lengths and angles at 300 K. The out-of-the-plane (apical) O2-Fe-O2 coordination is non-collinear for all the measured temperatures. So, there exists a local non centrosymmetry at $Fe^{3+}$ site.

For a better understanding of distortion, both the $Fe^{3+}$ and $Te^{6+}$ polyhedra are further subjected to the ellipsoid analysis (EA) [11]. The basis of the EA is the "minimum bounding ellipsoid" (MBE) fitting of polyhedra based on the "Khachiyan minimization algorithm" [22]. The algorithm is incorporated in a Python 2.7 based software package named PIEFACE (Polyhedra Inscribing Ellipsoids for Analyzing Crystallographic Environments) software package [11]. PIEFACE runs the algorithm to fit a set of data points in Cartesian coordinate within the set tolerance factor. The fitting can be initiated by the CIF (crystallographic information file) files obtained from the refinement of the diffraction pattern. Name (same as written in the CIF file) of the central cation in the polyhedral, maximum limit of radius, and tolerance factor is given as input control for the fitting. The resulting parameters are the three principal radii of the ellipsoid inscribing polyhedra ($R_1$, $R_2$, and $R_3$; $R_1 \geq R_2 \geq R_3$), rotation matrix of the ellipsoid, and off-centering displacement of the central cation with displacement vector. The resulting parameters are the three principal radii of the ellipsoid inscribing polyhedra ($R_1$, $R_2$, and $R_3$; $R_1 \geq R_2 \geq R_3$), rotation matrix of the ellipsoid, and off-centering displacement of the central cation with displacement vector. The size of the polyhedra is parameterized by the mean radii ($<R>$). The variance $\sigma^2(R)$ or standard deviation $\sigma(R) = \sqrt{\sigma^2(R)}$ defines the distortion. The ellipsoidal shape parameter is defined by, $S = \frac{R_3}{R_2} - \frac{R_2}{R_1}$. The parameters determine the polyhedra deformation in general, i.e., irrespective of the coordination, shape, and symmetry. Weak but finite off-centering displacement of $Fe^{3+}$ from the center of the inscribing ellipsoid is observed due to distortion in Fe polyhedra. Figure 3 (a) and (b) show the ellipsoidal fitting of different $Fe^{3+}$ and $Te^{6+}$ polyhedral at 300 K, respectively. Figure 4(i) shows the temperature variation of distortion and shape parameter $\sigma$ and $S$, respectively. In Fig. 4(ii), net displacement and average bond $<R>$ are shown. Prolate ($S>0$) type distortion is observed in $Fe^{3+}$ at 300 K, and the ion shifts along the negative $c$ direction. Sudden change in sign of S (from $S>0$ to $S<0$; i.e., prolate to oblate ellipsoid) is observed as the temperature is lowered. This sudden change in the shape parameter $S$ indicates the change in strain in the polyhedra [11]. Thus, the $Fe^{3+}$ site no longer behaves as an inversion center of the Fe polyhedra symmetry. The $S$ parameter and $\sigma$ parameter suggest non-octahedral type hybridization between Fe and O. The reduced magnetic moment of 4.35 $\mu_B$/$Fe^{3+}$ signifies the change in hybridization compared to regular octahedral coordination [23].

## B. The local structural study of $Fe^{3+}$ coordination polyhedra by X-ray absorption spectroscopy

The polyhedra distortion may be associated with the presence of mixed-valence state of $Fe^{2+}$ and $Fe^{3+}$ responsible for trimeron formation causing charge ordering induced ferroelectricity [24-26]. XAS investigation is performed to understand the charge state and coordination of Fe. XANE and EXAFS are sensitive local probes for charge state and ligand coordination around the central cation of polyhedra. Figure 5 shows Fe K-edge normalized XANES spectra of FTO sample at fixed temperature along with pure Fe metal, $Fe^{2+}$ and $Fe^{3+}$ standard. Fe K-edge XANES spectra confirm $Fe^{3+}$ valence state only. The inset of Fig. 5 shows background-subtracted pre-edges at different fixed temperatures. A sharp Gaussian-like pre-edge peak with low intensity is observed as expected for $Fe^{3+}$ having very small deviation from $D_{4h}$ symmetry. In regular $O_h$ coordination, a comparatively wide pre-edge with $e_g$ and $t_{2g}$ splitting is observed [27, 28]. A sudden fall in intensity near 210 K signifies the sudden change in the polyhedra shape as observed in the $S$ parameter from MBE analysis. Temperature variation of relative intensity ($I/I_{300\ K}$) is shown in the super-inset of the Fig. 5. The intensity of the pre-edge peak depends upon the hybridization in between the ligands and central cation in the polyhedra. The anomaly in the temperature variation of relative intensity $\sim T_N$ implies the correlation between magnetic order and Fe-O hybridization. The nonmonotonic temperature variation of intensity nonmonotonic variation of Fe-O bond-length or distortion as observed in the ND study.

For the analysis of the EXAFS data, the energy dependent absorption coefficient $\mu(E)$ has been converted to the energy dependent absorption function $\chi(E)$ and then to the wave number dependent absorption coefficient $\chi(k)$. Finally $k^2$ weighted $\chi(k)$ spectra was Fourier transformed in $R$ space to generate the $\chi(R)$ versus $R$ spectra in terms of the real distance from the center of the absorbing atoms. The analysis of the EXAFS data have been carried out following the standard procedures using the IFEFFIT software package, [29, 30] which includes Fourier transform (FT) to derive the $\chi(R)$ versus $R$ plots from the absorption spectra using ATHENA software, generation of the theoretical EXAFS spectra starting from an assumed crystallographic structure using ARTEMIS software [31] and finally fitting of $\chi(R)$ versus $R$ experimental data with the theoretical ones using the FEFF 6.0 code [32]. The Fe K-edge XAFS data is fitted considering FTO structure ($P4_2/mnm$). EXAFS data best fit (Fig. 6) were obtained to minimize $R_{factor}$ in the above process and defined by the below formula;

$$R_{factor} = \sum_i \frac{[Im(\chi_{dat}(r_i) - \chi_{th}(r_i))]^2 + [Re(\chi_{dat}(r_i) - \chi_{th}(r_i))]^2}{[Im(\chi_{dat}(r_i))]^2 + [Re(\chi_{dat}(r_i))]^2} \quad (3.1)$$

where, $\chi_{dat}$ and $\chi_{th}$ represent to the experimental and theoretical $\chi_R$ values respectively and *Im* and *Re* refer to the imaginary and real parts of the respective quantities. Bond lengths and corresponding Debye-Waller factors are enlisted in the Table I. Here, Fe is coordinated with two kinds of O atoms (Fe-O2; apical position) at 1.995(3) Å distance and four O atoms (*Fe-O1/O2*; in-plane position) at 2.032(3) Å distance in the first shell. The next nearest-neighbors are considered at various distances: two *Fe/Te* atoms (*Fe-Fe1/Te1*) at 2.645(3) Å, six oxygen atoms (*Fe-O3*) at 3.795(3) Å, four *Te* atoms (Fe-Te2) at 3.728(4) Å, and four Fe atoms (*Fe-Fe2*) at 3.525(4) Å at RT. The fittings were performed in phase un-corrected R-space range of 1-3.6 Å. The $Fe^{3+}$ first shell modeling as approximated $D_{4h}$ polyhedra is consistent with ND and XRD analysis. The in-plane Debye-Waller factor corresponding to the equatorial *Fe-O* bonding is 0.0060(3), suggesting a deviation of $Fe^{3+}$ from $O_h$ symmetry. *Fe-O2* (3.795 Å) bonding connects the $Fe^{3+}$ polyhedra on the *c*-axis and sub-compartmental polyhedra through corner-sharing. Corresponding large Debye-Waller factor 0.0073 (2) suggest larger flexibility along with the bonding [33]. Thus XAS study concludes about the $Fe^{3+}$ charge state and distorted Fe-O coordination locally.

## C. Charge density distribution calculation by MEM analysis

The off-centering of $Fe^{3+}$ with polyhedra distortion can be inferred by ND and XAS analysis to suspect FE polarization at RT. FE polarization is also induced due to asymmetric charge density distribution caused by asymmetric hybridization in presence of magnetic correlation [34]. Thus, electronic contribution should be explored to explain the origin of FE if any above $T_N$. To understand the electronic contribution, the electron density (ED) distribution is calculated at 300 K employing maximum entropy method (MEM) [35, 36] analysis using the DYSNOMIA program [37].

Rietveld refinement of the synchrotron X-ray diffraction data is done to obtain the structure factor for the initiation of charge density distribution calculation. Figure 7(a) shows the Rietveld refined of synchrotron XRD pattern with reliability parameter of the fitting in the inset. The Rietveld refinement and input file (.mem) containing the structure factor and other instructions are processed in FULLPROF software [21] to calculate density following the maximum entropy method (MEM). The calculation is initialized with the structure factor obtained from the Rietveld refinement of high energy synchrotron XRD (λ=0.6304Å) data. The voxels dimension is 128×128×128 in the unit cell to calculate the 3d charge density map over the unit cell. Figure 7(b) shows the 3-dimensional charge density distribution map along with the crystal structure obtained from the Rietveld refinement. The perfect overlapping of the calculated 3-dimensional charge distribution with the structure obtained from the Rietveld refinement is consistent with each other. This validates the calculation of charge density distribution. Figure 7 shows the 2d ED contour map on (1-10) plane inside 3d unit cell and separately (Fig 7(b)) to understand the covalent electron density around different ionic core sitting at the center of polyhedra. The contours are plotted in the range 0 to 1 with step size 0.1. The edge-sharing of iron octahedral and 2d density profile along [001] and $\left[\frac{1}{2}\frac{1}{2}1\right]$ direction can be realized by the plot. The density distribution on either side of the $Fe^{3+}$ on the equatorial plane of the polyhedra is found to be asymmetric. The ion core of $Fe^{3+}$ and O1 and O2 are not spherically symmetric, and distorted charge density is observed around the ionic center of these cores. The $Te^{6+}$ ion core and the density distribution around it are found to be symmetric. The $Te^{6+}$ ion core and the density distribution around it are found to be symmetric. The line density profile (Fig. 8) is obtained for equatorial in-plane metal-ligand bond Fe-O1, Fe-O2, and also for apical Fe-O2. Covalent electron density and midpoints (saddle point of covalent Fe-O bonding) along the bonds show different variations for each case. Midpoints and midpoint electron density from Fe end in Fe-O1 (equatorial) are 0.921 Å and 0.420 e/Å$^3$, respectively and for Fe-O2 are 0.951 Å and 0.555 e/Å$^3$ respectively. Whereas, midpoint and electron density of the out of plane Fe-O2 (apical) is 0.941 Å and 0.422 e/Å$^3$, respectively. Thus, ferroelectric polarization may exist due to asymmetric electron density around $Fe^{3+}$. Microscopic polarization is estimated from the shift (*Δc*) of each ionic center and charge cloud center for all cations and anions present in the unit cell, as partially shown in Fig. 9. The position of the ionic center is obtained from the Rietveld refinement of synchrotron XRD data, and the center of the charge cloud is obtained from the center of the minimum charge density contour. The polarization (*P*) is calculated using the formula, $P = \frac{ne\Delta c}{V}$; *n* is the number of electrons, *e* is the electronic charge, and *V* is the volume of the unit cell. The net polarization value is P~0.05μC/cm$^2$, obtained by adding individual ionic contributions. Hence, we can conclude that the FE at 300 K is mainly driven by off-centering of *Fe3+* and electron density distortion around other ionic centers.

**D. Magnetoelectric coupling in FTO:**

The detailed study of ME coupling and its correlation with structural and magnetic properties are addressed via different approaches like magnetic field dependent FE (PE loop) and magnetic field dependent ME voltage study. The presence of magnetoelastic coupling is further studied by temperature and magnetic field-dependent strain measurement.

i. **Magnetic field dependent PE loop measurement:**

The presence of room temperature FE and ME coupling is observed by direct PE hysteresis loop measurement using the special remanent protocol at room temperature [38, 39]. The FE hysteresis loop with remanent polarization value ~5 nC/cm$^2$ is observed, as shown in the upper inset of Fig. 9(a). The variation of the PE loop is further studied under different fixed magnetic fields (0 T-1.2 T). A conspicuous change in remanent polarization is observed due to the applied magnetic field. This is a direct confirmation of the presence of FE and ME coupling in FTO at room temperature. The degree of coupling is parameterized by $MP_r = \frac{P_r(H) - P_r(0)}{P_r(0)}\%$ to understand the field variation of the coupling as shown in the lower inset of Fig. 9(a).

The origin of FE at 300 K and the ME coupling above $T_N$ can have several possibilities origins as follows; i) FE induced trimeron formation [24-26] due to charge ordering of multiple charge state of Fe as $Fe^{2+}$ and $Fe^{3+}$ but $Fe^{3+}$ charge state is confirmed by the XAS study, ii) Symmetry breaking due to nonrelativistic exchange striction at AFM domain wall boundary as seen in some collinear antiferromagnets [39], and iii) Weak off-centering of cation and/or asymmetric charge distribution although the noncentrosymmetric polar phase is nor observed globally [40]. The reason ii) can be discarded since the existence of the magnetic domain at 300 K is not possible as any kind of long-range order could not be observed at this temperature. Thus, the origin of FE in the material can be interpreted in terms of the off-centering of $Fe^{3+}$ and the asymmetric charge density distribution of the valence electrons around different ion cores as elaborated in the structural study via ND, XAS, and charge density distribution analysis. The distorted Fe polyhedra share edges with Te polyhedra on one side and Fe polyhedra on another side (Fig. 2(a)). This edge-sharing is doubly bridged by oxygen on both the sides, which interconnects the polarization and magnetic interaction. The significant change in remanent polarization on the application of the magnetic field can be addressed by the weak/short-ranged magnetic correlation present in the material above $T_N$ [16-18]. ME coupling above $T_N$ is mainly driven by the paramagnetic contribution along with the weak magnetic contribution due to the mutual interaction of $J_1$ and $J_{dim}$ between two nearest neighbor $Fe^{3+}$ as shown in Fig. 2(a).

### ii. ME voltage measurement:

ME coupling is also addressed by measuring the ME voltage following the dynamic method [41]. The operating frequency (199.9 Hz) of the measurement is chosen to minimize the phase shift between the applied ac signal and the measured voltage. The in-phase ME voltage difference as a function of the applied dc biasing magnetic field at a different fixed temperature is shown in Fig. 10(a). Nonlinear variation of ME voltage difference with the signature of negative ME coupling is observed at 300 K. The nonlinear ME response at 300 K signifies the presence of interdependent polarization and magnetic correlation at room temperature as the shreds of evidence were also seen in structural, magnetization, specific heat, and μSR studies.

As the temperature decreases, the gradual increase in the Fe-O2-Fe superexchange (J2) strength drives the system to the long-range ordered state. The different exchange pathway responsible for the ME coupling is schematically shown in Fig. 2(a). The ME voltage responses follow the magnetic ordering, and the presence of a mixed interaction ($J_1+J_{dim}$ and $J_2$) is replicated in the ME responses, as seen in Fig. 10(b). Negative nonlinear behavior is disrupted, and a complex ME voltage response is evolved in the temperature range 210 K-300 K. A clear peak at 0.5 T is anticipated as the temperature decreases through $T_N$. The negative response gradually changes to an irreversible nonlinear variation with the applied bipolar dc magnetic field (±1.2 T) below $T_N$. In this temperature regime, the spin-phonon coupling is initiated as the long-range AFM ordering builds up due to stronger AFM correlation via Fe-O2-Fe pathway. The spin-phonon coupling influences the ME coupling below $T_N$ via Fe-O2-Fe pathway. In the long-range ordered state, two competing interactions (Fig 1(a)); namely, i) intrabilayer interaction having exchange strength $J_1+J_{dim}$ and ii) inter bilayer interaction having exchange strength $J_2$, result in such kind of a complex ME voltage response. Below 0.5 T, the ME response is dominantly contributed by the interbilayer interaction, whereas intrabilayer contribution plays a significant role above 0.5 T. The interbilayer magnetic correlation is not established at room temperature, and the intrabilayer correlation manifested due to the $Fe^{3+}$ - $Fe^{3+}$ $J_{dim}$ and $J_1$ is predominantly active. Thus, a strong nonlinear ME response is observed instead of a linear negative response commonly, observed due to paramagnetic contribution well above the magnetically ordered state. The nonlinear field dependence of the ME voltage suggests the presence of magnetoelasticity in the material. Evidence of magnetoelastity and its role in describing the nonlinear ME coupling is discussed as follows.

### iii. Temperature and magnetic field dependent strain measurement:

The nonlinear behavior of the ME voltage suggest strong coupling between the spin and lattice degrees of freedom. The presence of magnetoelastic coupling is confirmed by temperature and magnetic field-dependent strain measurements. The temperature-dependent strain is recorded by measuring the temperature dependence of the resistance of the strain gauge alone and also by sticking the gauge on the sample. The difference between the gauge contribution from the sample and the gauge gives the relative change in the resistance. This relative change in the resistance is then converted into strain using the gauge factor. The measurement protocol thus minimizes the contribution of the strain gauge, and the sample contribution is maximally extracted. The temperature-dependent strain is measured at 0 T, 0.6 T, and 1.2 T by cooling down the sample to the lowest temperature then the measurement is done during the warming. The magnetic field is applied at the lowest temperature before the warming for temperature-dependent measurement at fixed magnetic fields. The derivative of the temperature-dependent strain variation is shown in Fig. 11(a). An apparent anomaly can be seen near the magnetic transition temperature at ~210 K. The temperature-dependent strain is shown in the inset of Fig.

9(a). The strain increases more rapidly below 210 K. The derivative of the zero-field data clarifies the anomalies consistent with the magnetic transition. Temperature variation of the magnetostrain $\left(\frac{\varepsilon(H)-\varepsilon(0)}{\varepsilon(H)}\right) \times 100$ is shown in Fig. 11(b). Two distinct features appear at ~$T_N$, and ~250 K signifies the correlation of magnetostrain with magnetic ordering. The magnetic field variation of strain at 120 K is shown in the inset of Fig. 10(b). It can be seen that the field variation in magnetostrain changes around the magnetic transition. At 100 K, a robust nonlinear field dependency of the magnetostrain is observed in comparison to the RT where the effect is found to be insignificant. The magnetoelastic effect is dominantly imprinted as the long-range order gradually builds up due to strengthening the inter-bilayer exchange $J_2$ via Fe-O2-Fe exchange pathway. The effect is indirectly observed via spin-phonon coupling. The anisotropic behavior of the magnetoelastic coupling is responsible for such kind of ME voltage and magnetostrain response.

### IV. Summary:

In summary, we review the structural, magnetic, and ME properties of the inverse trirutile $Fe_2TeO_6$ compound. Thereafter, the following unaddressed interesting structural, magnetic, and ME correlation in the material is concluded. The presence of noncentrosymmetry at the Fe site due to off-centering displacement of $Fe^{3+}$ is revealed. The asymmetric charge density distribution of covalence electrons is also found in $Fe^{3+}$ polyhedra. This local distortion in the Fe polyhedra is manifested due to the asymmetric hybridization between central cation and ligand.

Direct evidence of intrinsic ME coupling is presented by showing intercoupled spin-charge-lattice above and below $T_N$. The unconventional existence of magnetic field-dependent FE and its possible origins are critically discussed to conclude the involvement of both lattice (off-centering of $Fe^{3+}$) and electronic distortion (asymmetric charge density distribution of covalence electron) as manifested by detail structural investigations. ME coupling above $T_N$ is mediated by intrabilayer exchange. Below $T_N$, ME coupling is mediated by the interbilayer exchange, which dominantly determines the nature of ME coupling along with intrabilayer exchange. Hence, both the FE and ME coupling solely depend upon the $Fe^{3+}$, though we observe ME coupling above $T_N$. The nonlinear hysteretic nature of ME coupling suggests the presence of magnetoelastic coupling, which is concluded by magnetic field-dependent macroscopic strain measurements below $T_N$. Thus the origin of FE and the unconventional nature of ME coupling are discussed thoroughly, conveying the unsaturated part of the motivation to study $Fe_2TeO_6$ and other trirutile antiferromagnets in the future. The material can be further studied for optimization towards magnetic sensor application by exploiting the observed ME and magnetostrain properties.

**Acknowledgment:** We acknowledge UGC-DAE-CSR Mumbai, India, for the project grant (Sanction No. CRS-M-187, 225). We acknowledge Arkadev Roy, BL-11 of Indus-2 synchrotron source at the Raja Ramanna Centre for Advanced Technology (RRCAT)) Indore, India. We acknowledge Dr. Dipten Bhattacharya and Aditi Sahoo, CSIR CGCRI, Kolkata, for PE loop measurement.

Figure Captions:

Figure 1: The Rietveld refined fitted pattern at various fixed temperatures with reliability parameters. The black circular symbols denote the observed intensity, while the solid red line represents the calculated pattern. The green bars represent the nuclear and magnetic Braggs peak. The blue lines below represent the difference between the calculated and observed diffraction patterns.

Figure 2. (a) Crystal and magnetic structure obtained from the Rietveld refinement of ND data at 3K. The side panel shows the distorted $Fe^{3+}$ polyhedral. (b) Temperature variation of lattice parameters ($a$, $c$), and $Fe^{3+}$ moment (inset).

Figure 3. Minimum bounding ellipsoid (MBE) fit to $Fe^{3+}$ (a) and $Te^{6+}$ (b) polyhedra at (a) 300 K, (. The left side of each of the figures shows the different parameters; center of the ellipsoid, off-centering displacement, average radius, polyhedral distortion, the volume of the ellipsoid, shape parameter, and the fitting error.

Figure 4. Temperature variation of (i) polyhedra distortion *(σ(R))*, shape *(S)*; (ii) average bond length *(<R>)*, and the off-centering displacement (*D*) of $Fe^{3+}$ as obtained from the ellipsoid analysis.

Figure 5. Normalized Fe K-edge XANES spectra of FTO sample as a function of temperature along with pure Fe metal, $Fe^{2+}$, and $Fe^{3+}$ standard. The inset shows pre-edge peak variation at various fixed temperatures. The superinset shows the temperature variation of the relative intensity ($I/I_{300\ K}$) of background-corrected pre-edge peak of XANES spectra.

Figure 6. $k^2$-weighted spectra of Fe K-edge XAFS; (i) modulus of the [|χ(R)|] and (ii) real part of the [Re|χ(R)|] at various temperatures. The symbol shows data, and solid lines are the best fitting.

Figure 7. (a) The Rietveld refined fitted pattern of synchrotron X-ray diffraction data taken at 300 K. The black circular symbols denote the observed intensity while the solid red line represents the calculated pattern. The green bars represent the Braggs peak. The blue line below represents the difference between the calculated and observed diffraction patterns. (b) Three-dimensional electron density distributions map with lattice structure. The inside 2d plane refers to the electron density distribution on the 1-1 0 planes.

Figure 8. 2*d* electron density plot on 1 -1 0 planes to show the charge density distribution around $Fe^{3+}$ and $Te^{6+}$

Figure 9. Different 1$d$ ligand bond (Fe-O) density distribution in $Fe^{3+}$ polyhedra as obtained by charge density distribution calculation by MEM analysis in Dysnomia. The figure confirms the asymmetric charge density along with different Fe-O ligation and thus distorted $Fe^{3+}$ Polyhedra. (b) Some of the positions of the ionic centers (IC) and the charge cloud centers (CC) show non-overlapping and possibility of ferroelectric polarization.

Figure. 10. Magnetic field variation of Magnetoelectric voltage difference at (a) 300 K. Upper inset: Polarization (P) versus electric field (E) at different fixed magnetic fields and lower inset: $MP_r\%$ versus magnetic field, (b) 250 K, (c) 225 K, and (d) 125 K.

Figure. 11 (a) Temperature derivative $\left(\frac{d\varepsilon}{dT}\right)$ of the strain versus temperature. Inset shows zero-field temperature-dependent strain ($\varepsilon$). (b) The temperature dependency of magnetostrain $\left(\frac{\Delta\varepsilon}{\varepsilon}\%\right)$ at 0.6 T and 1.2 T. Inset: Magnetic field variation of $\varepsilon$ at 120 K.

**TABLE I**. Structural parameters obtained from Fe K-edge XAFS fitting are the following: CN (co-ordination number), R (bond distance), and $\sigma^2$ (Debye-Waller factor). The numbers in parentheses indicate the uncertainty in the last digit.

| | | RT | | 210 K | | 180 K | | 15K | |
|---|---|---|---|---|---|---|---|---|---|
| Shell Path | N | R(Å) | $\sigma^2$ (Å$^2$) | R(Å) | $\sigma^2$ (Å$^2$) | R(Å) | $\sigma^2$ (Å$^2$) | R(Å) | $\sigma^2$ (Å$^2$) |
| Fe-O1 | 2 | 1.995(3) | 0.0042(2) | 1.994(3) | 0.0042(3) | 2.003(3) | 0.0039(3) | 2.005(3) | 0.0036(3) |
| Fe-O2 | 4 | 2.032(3) | 0.0060(3) | 2.023(3) | 0.0058(4) | 2.017(3) | 0.0054(3) | 2.016(3) | 0.0052(3) |
| Fe-Fe1/Te1 | 2 | 2.645(3) | 0.0127(4) | 2.629(3) | 0.0103(3) | 2.639(4) | 0.0129(3) | 2.646(4) | 0.0118(3) |
| Fe-O2 | 6 | 3.795(3) | 0.0073(2) | 3.777(4) | 0.0073(3) | 3.762(3) | 0.0073(4) | 3.777(3) | 0.0065(3) |
| Fe-Te2 | 4 | 3.728(4) | 0.0059(4) | 3.713(3) | 0.0057(4) | 3.703(3) | 0.0046(3) | 3.712(3) | 0.0037(3) |
| Fe-Fe2 | 4 | 3.525(4) | 0.0091(3) | 3.525(3) | 0.0094(3) | 3.513(4) | 0.0087(4) | 3.509(4) | 0.0084(3) |

Fig. 1

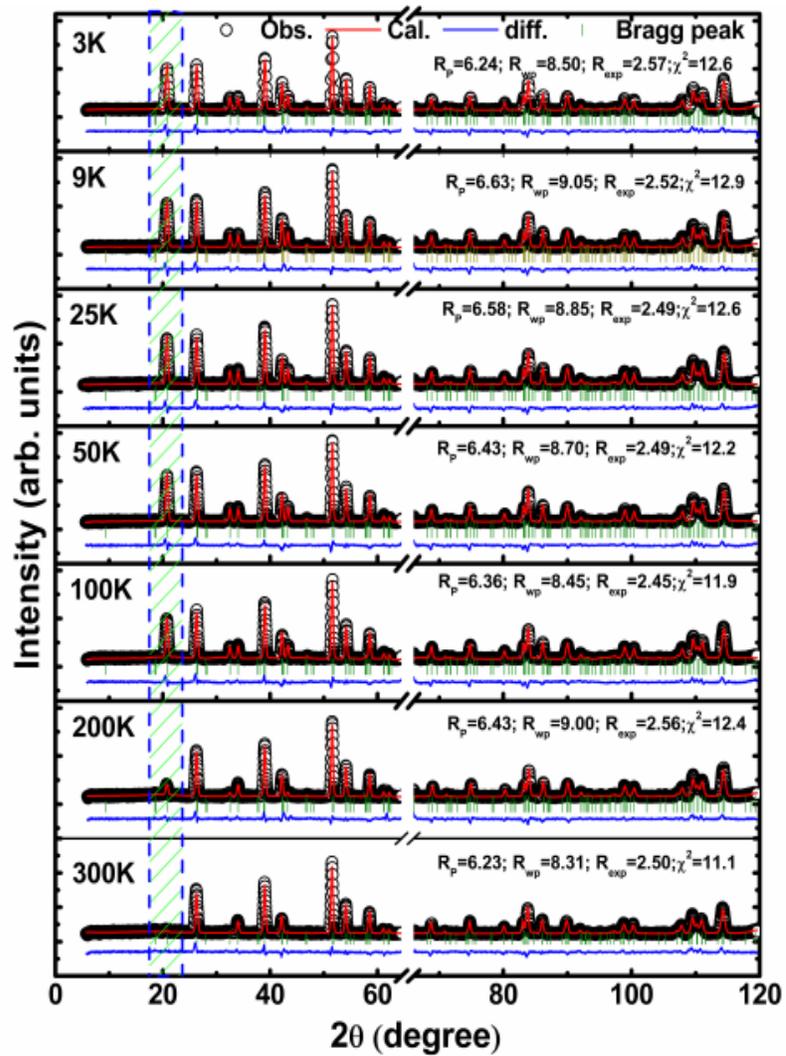

Fig 2.

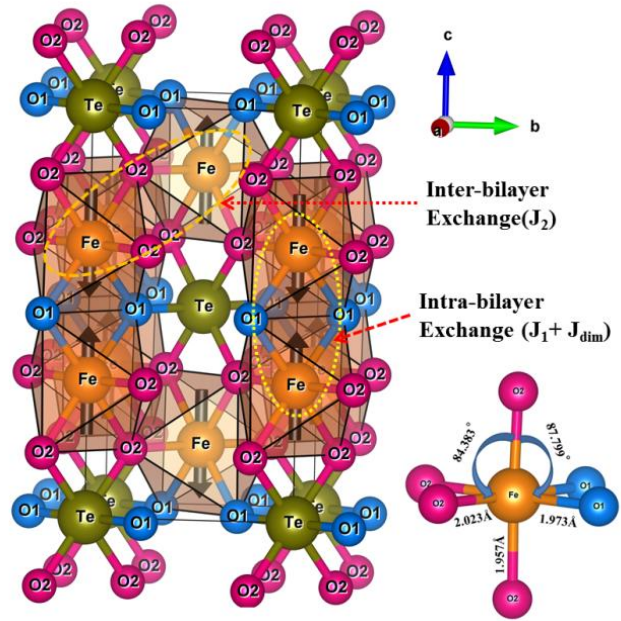

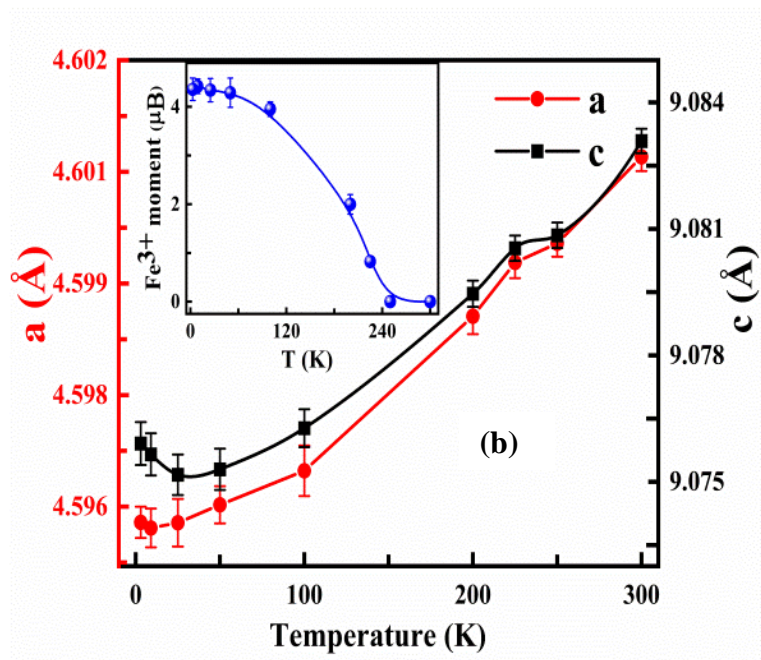

Fig. 3.

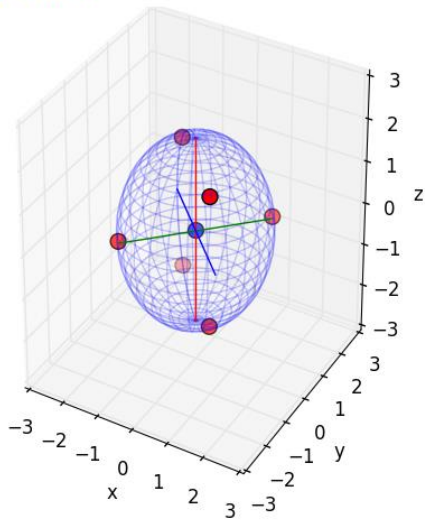

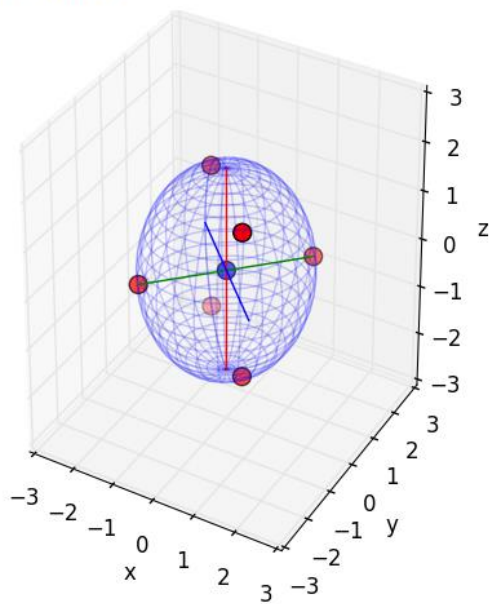

Fig. 4

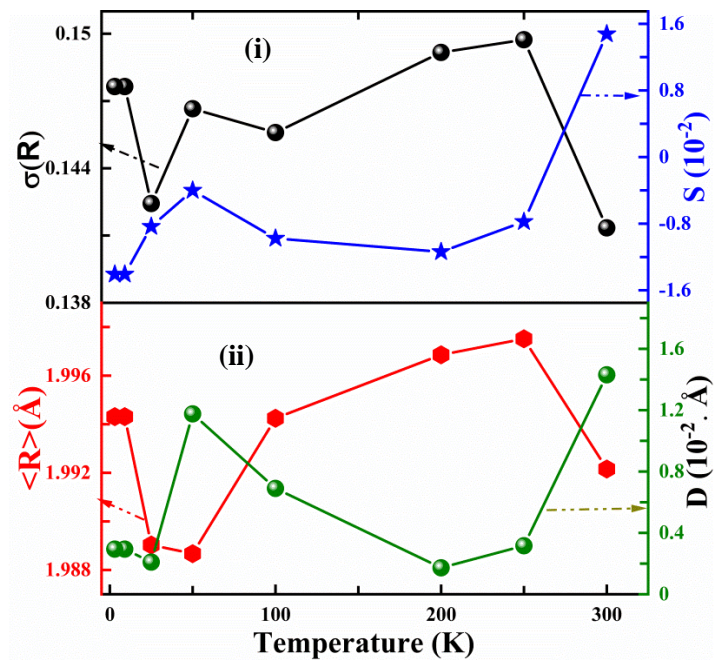

Fig. 5.

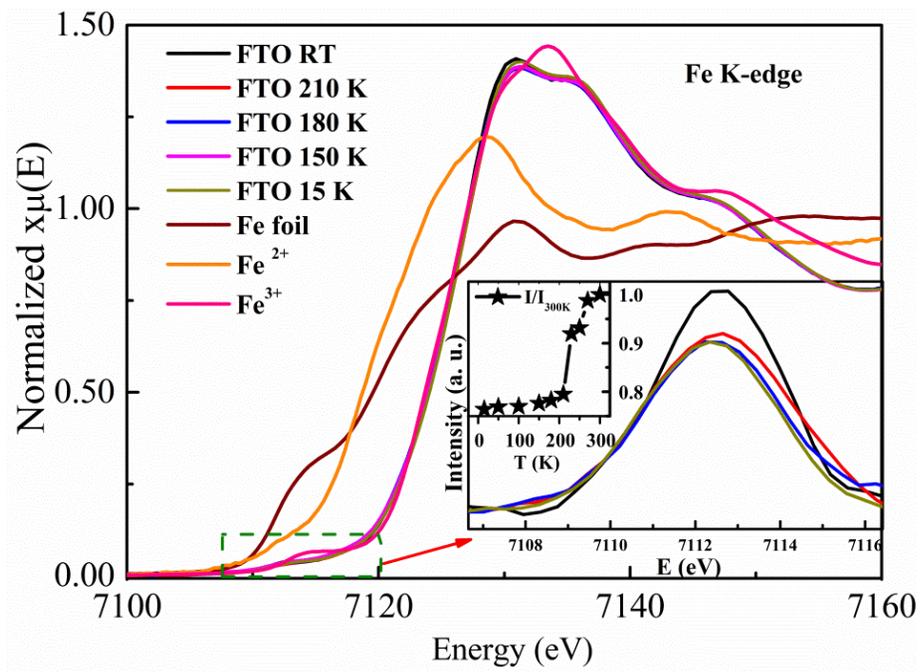

Fig. 6.

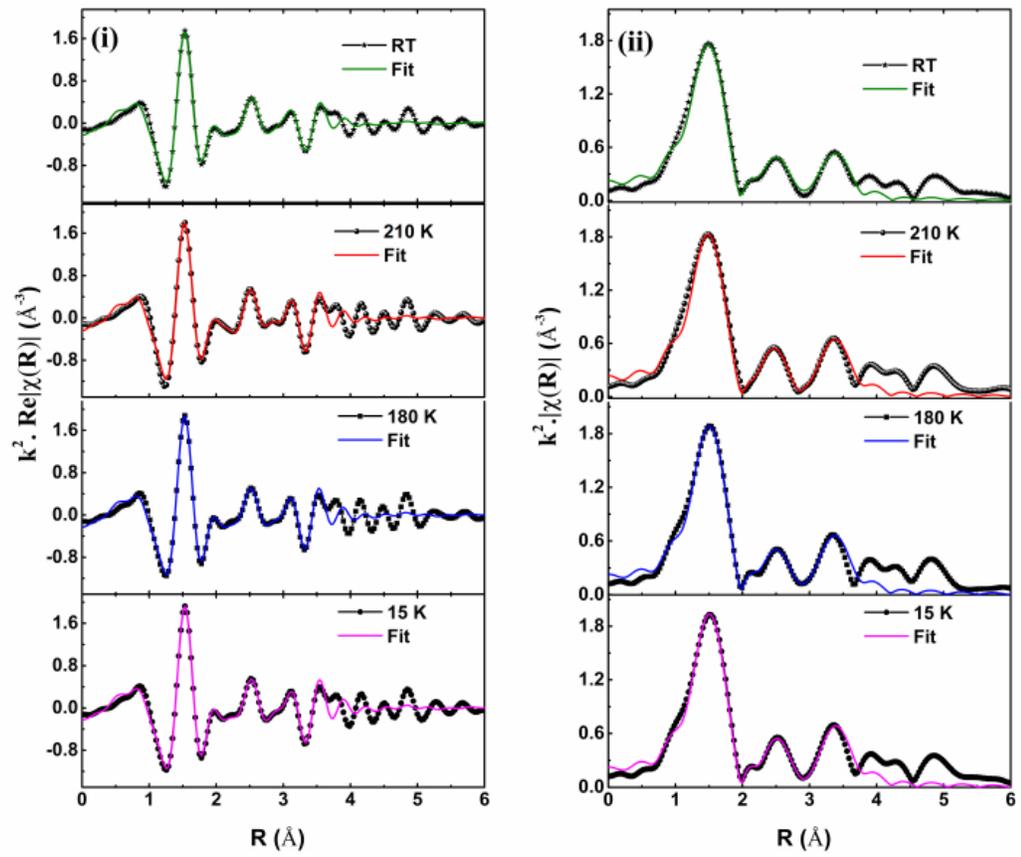

Fig. 7.

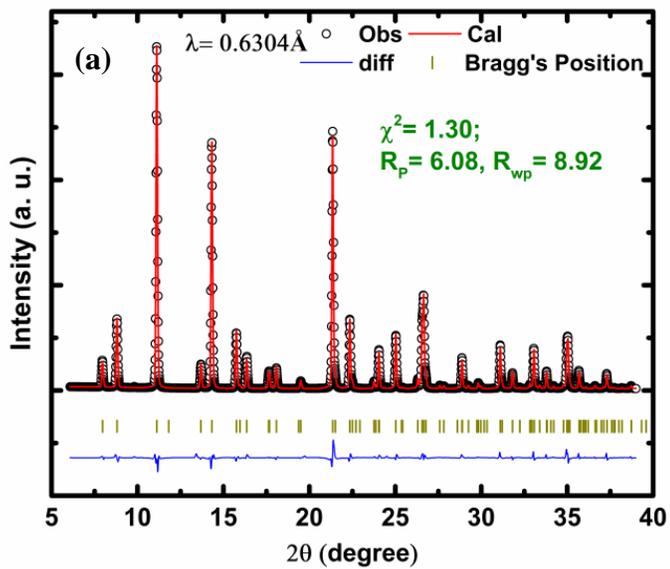

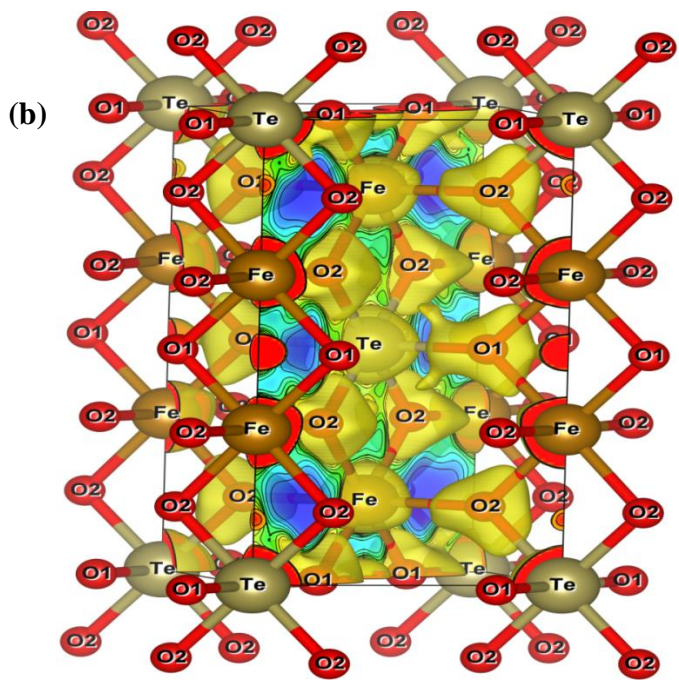

Fig. 8

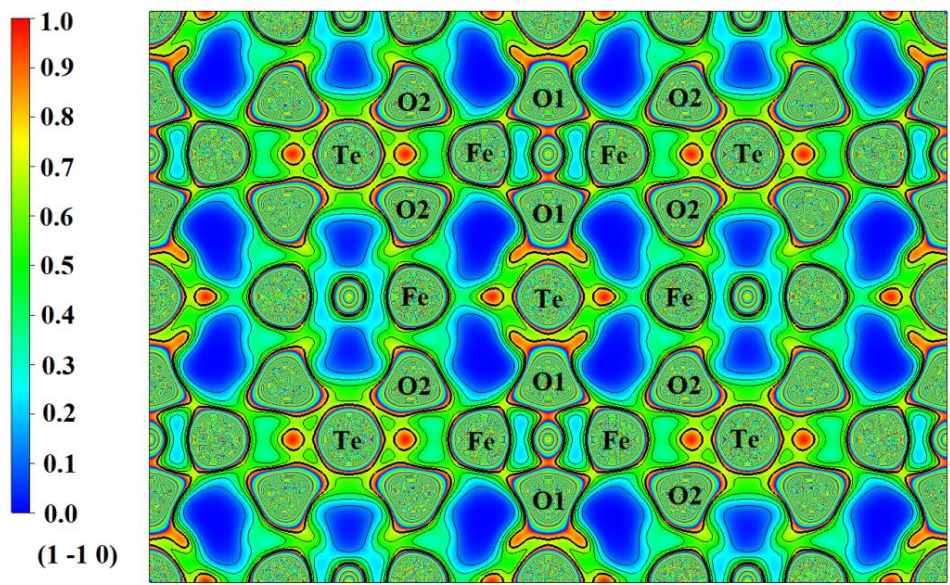

Fig. 9

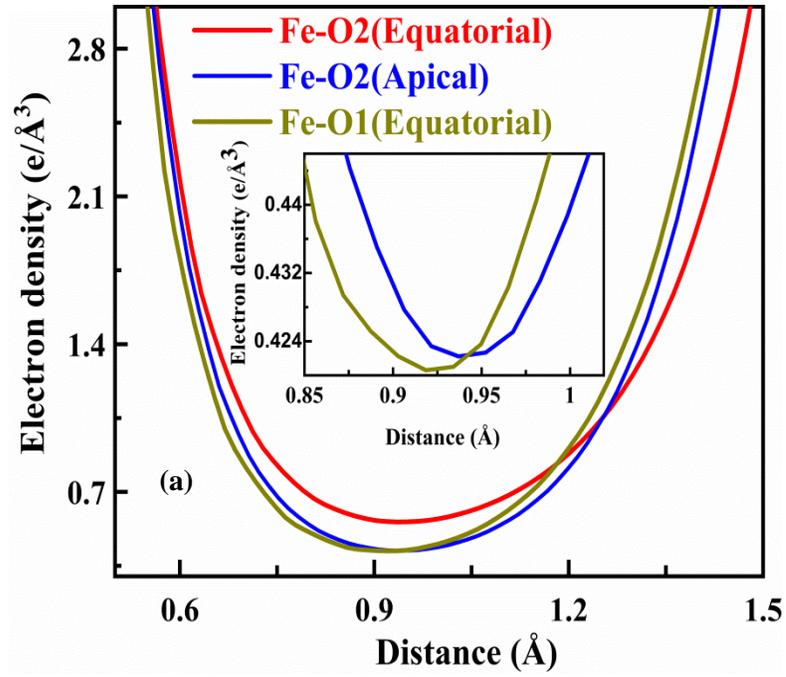

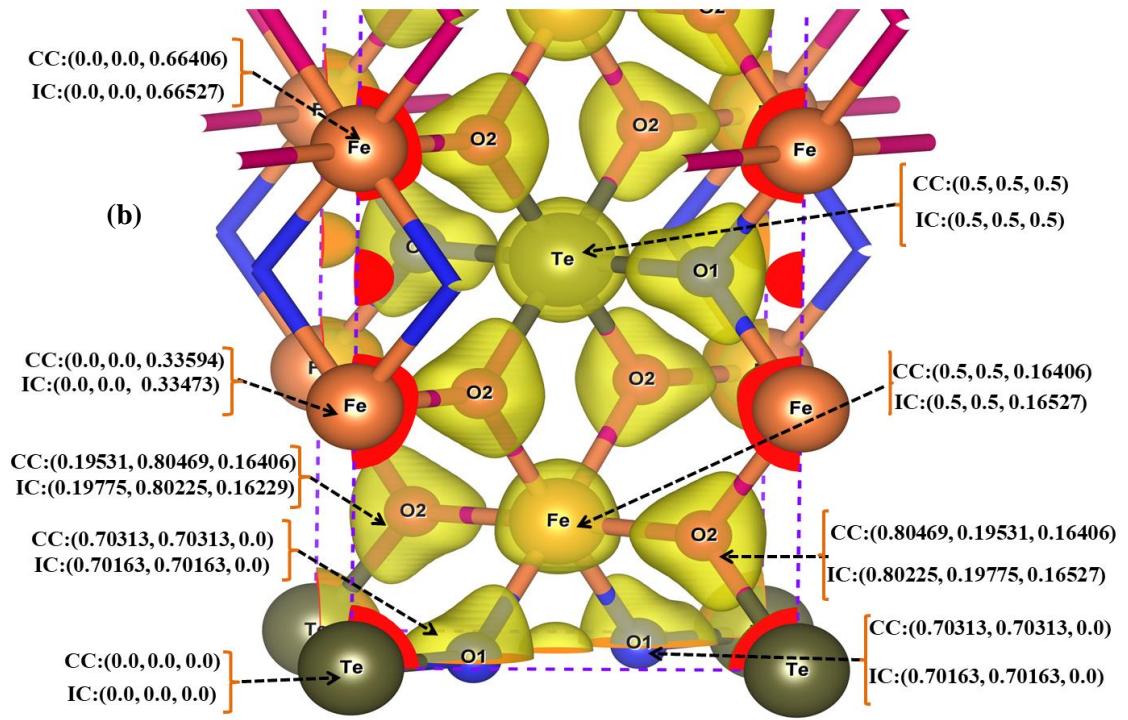

Fig. 10

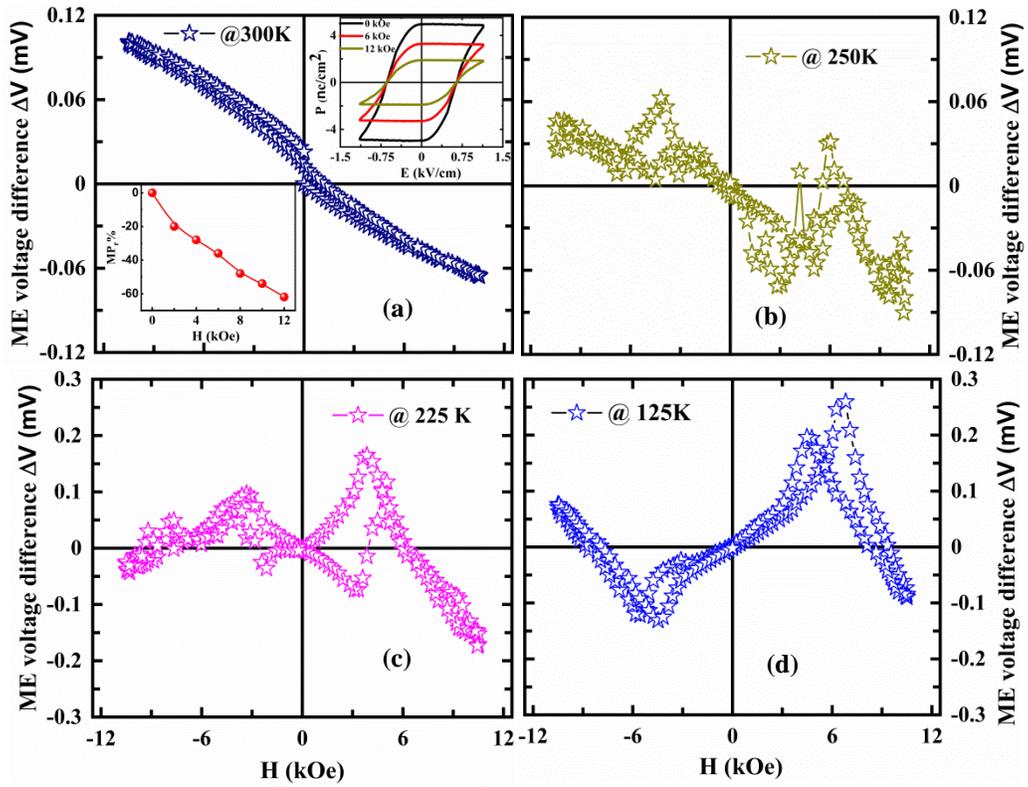

Figure 11.

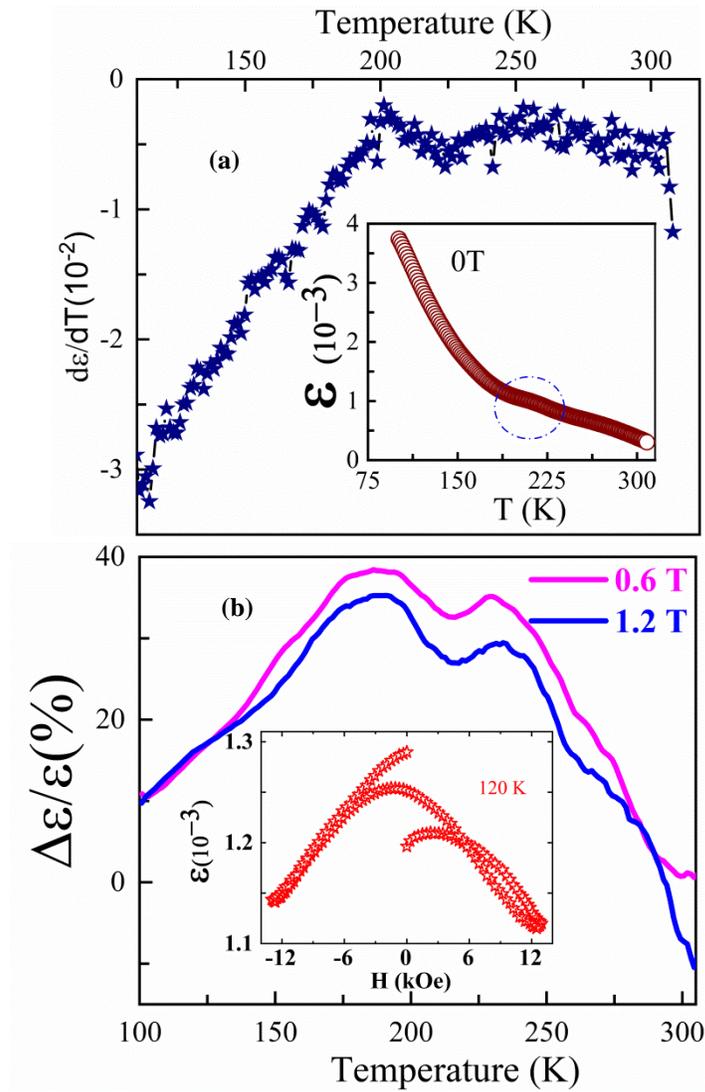